\documentclass[12pt]{iopart}
\usepackage{graphicx} 

\def\prd{{\em Phys. Rev. {\bf D} }}

\def\npb{{\em Nucl. Phys. {\bf B} }}
\def\epjc{{\em Eur. Phys. J. {\bf C} }}

\begin{document}

\title{ Evolution of mechanism of parton energy loss with transverse momentum
at RHIC and LHC in relativistic collision of heavy nuclei}
\title[Evolution of parton energy loss...]{}

\author{Dinesh K. Srivastava}
\address{Variable Energy Cyclotron Centre, 1/AF Bidhan Nagar, Kolkata 700064, 
India}

\begin{abstract}

We analyze the suppression of particle production at large transverse
momenta in ($0-5\%$ most) central collisions of gold nuclei at
$\sqrt{s_\textrm{NN}}=$ 200 GeV
and lead nuclei at
 $\sqrt{s_{\textrm{NN}}}=$ 2.76
 TeV. Full next-to-leading order radiative corrections at
${\cal{O}}(\alpha_s^3)$,
and nuclear effects like shadowing and parton energy loss are included.
The parton energy loss is implemented in a simple
 multiple scattering model, where the partons
lose an energy $\epsilon=\lambda \times dE/dx$ per collision,
where $\lambda$ is their mean free path.  We take $\epsilon=\kappa E$
for a treatment which is suggestive of the Bethe Heitler (BH) mechanism
of incoherent scatterings,
$\epsilon = \sqrt{\alpha E}$ for LPM mechanism, and  $\epsilon=$
constant for a mechanism which suggests that the rate of energy loss
($dE/dx$) of the partons is proportional to total path length 
($L$) of the parton in the plasma, as the formation time of the radiated
gluon becomes much larger than $L$.
 We find that while the BH mechanism describes the
nuclear modification factor $R_{\textrm{AA}}$ for $p_T \leq$ 5 GeV/$c$ 
(especially at RHIC energy), the LPM
and more so the constant $dE/dx$ mechanism provides a good description at
larger $p_T$. This confirms the earlier expectation that the energy loss
mechanism for partons changes from BH to LPM for $p_T \ge \lambda <k_T^2>$,
where $\lambda \approx$ 1 fm and $<k_T^2> \approx$ 1 GeV$^2$ is the average
transverse kick-squared received by the parton per collision.
The energy loss per collision at the $\sqrt{s_\textrm{NN}}$ =2.76 TeV
is found to be  about twice of that at 0.2 TeV.

\end{abstract}

\section{Introduction}

The suppressed production of particles having large transverse momenta,
known as jet-quenching~\cite{jetq_theo,jetq_exp} 
along with the
strong collective flow \cite{v2_exp,v2_theo} and the success of recombination
model~\cite{recomb} in explaining the constituent quark number scaling
 observed for the elliptic flow, at the experiments performed at
the Relativistic Heavy-Ion Collider (RHIC)
at Brookhaven National Laboratory have provided
a strong evidence for the production
of quark gluon plasma (QGP) in collisions of nuclei at relativistic energies.
The jet-quenching~\cite{jet_lhc} and elliptic flow~\cite{v2_lhc} have already
been confirmed in collision of lead nuclei at $\sqrt{s_{\textrm{NN}}}=$ 2.76 TeV
at the Large Hadron Collider.
\begin{figure}[ht]
\begin{center}
\includegraphics[width=30pc,angle=-0]{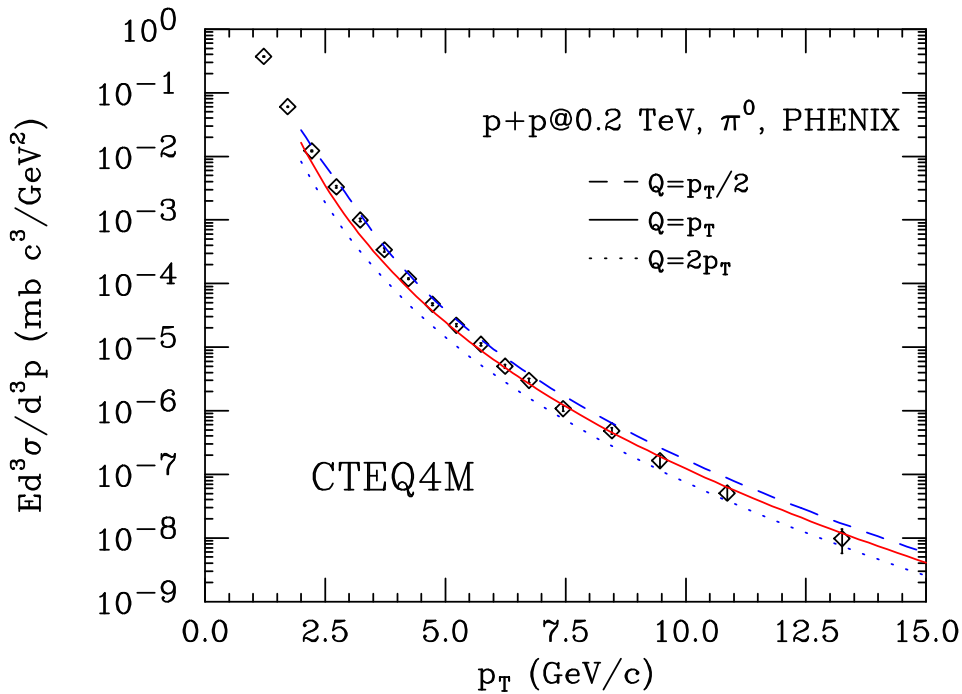}
\end{center}
\caption{A comparison of inclusive production of neutral
 pions~\cite{phenix_pp}
with NLO pQCD calculations at $\sqrt{s_\textrm{NN}}$ = 200 GeV.}
\label{fig1}
\end{figure}
The simplest measure of the jet-quenching at large $p_T$ is given by
the nuclear modification factor $R_{\textrm{AA}}$, which provides the ratio
of particle yield in AA collisions to that in pp, defined as:
\begin{equation}
R_{\mathrm{AA}}=\frac{(1/N^{\mathrm{AA}}_{\mathrm{evt}})\,
d^2N_{\mathrm{AA}}/d^2p_Tdy}
{ {N_{\mathrm{coll}}\, (1/N^{\mathrm{pp}}_{\mathrm{evt}}})\,
d^2N_{\mathrm{pp}}/d^2p_Tdy}
\end{equation}
where $N_{\textrm{evt}}$ is the number of events and $N_{\textrm{coll}}$ is the
number of binary (nucleon-nucleon) collisions. In absence of any
nuclear modification, $R_{\textrm{AA}}$ would be unity at large $p_T$.
  The larger
density of the medium likely at the higher LHC energies 
is expected to give rise to a larger loss of energy for a given  $p_T$,
 which should
lead to a larger suppression. Yet the expectation that the spectrum of the
partons at large $p_T$ will fall less steeply at the LHC would mean a smaller
suppression for the same energy loss of parton of a given $p_T$. Thus
indeed the $R_{\textrm{AA}}$ at LHC shows an interesting behaviour: it decreases
with increase in $p_T$ in the window of $p_T \approx$ 2--5 GeV/$c$ and then
rises to reach a value of about 0.4 at $p_T\approx $ 20 GeV/$c$. The 
results for the 
most central collisions at RHIC energies can also be interpreted
to suggest the emergence of
a similar trend. In the present work we suggest that this result is indicative
of change of the mechanism for energy loss as $p_T$ rises beyond 5--7 GeV/$c$
as the Landau Pomeranchuk Midgal effect sets in to suppress radiation of
gluons.

In the next section we give the formulation of our treatment and introduce
various mechanisms of energy-loss used in our work. In section III, we discuss
our results at RHIC and LHC energies for the 0--5\% most central collisions.
Finally we give a short discussion and conclusions.
\begin{figure}[ht]
\begin{center}
\includegraphics[width=30pc,angle=-0]{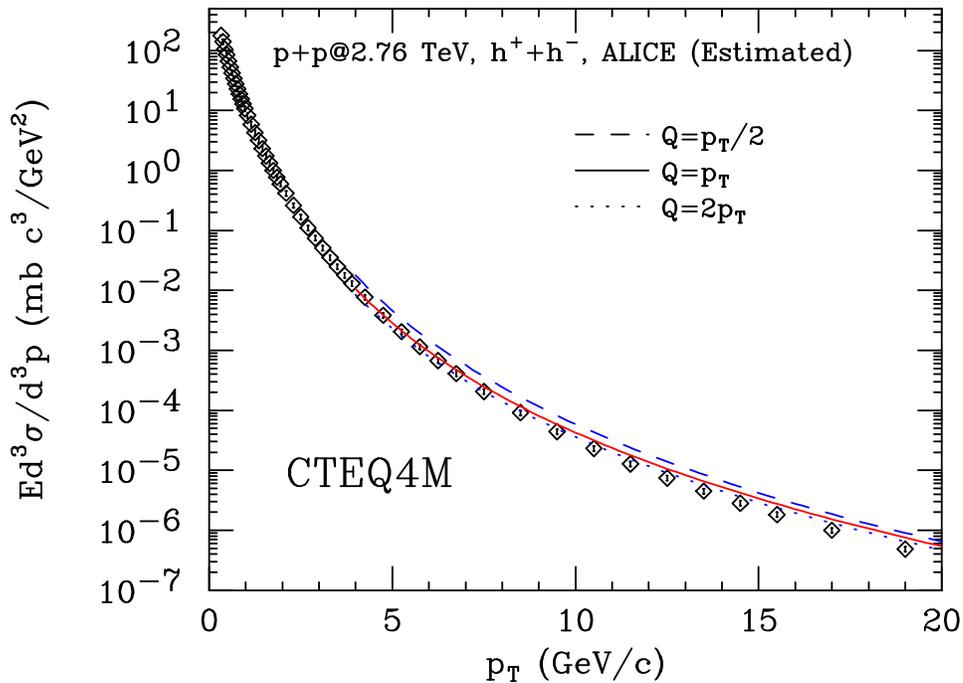}
\end{center}
\caption{A comparison of inclusive production of charged hadrons
estimated by the ALICE collaboration~\cite{jet_lhc}
with NLO pQCD calculations at $\sqrt{s_\textrm{NN}}$ = 2.76 TeV.}
\label{fig2}
\end{figure}
\section{Formulation}
\subsection{Particle production in $pp$ collisions}
As a first step we check if the next-to-leading order perturbative 
QCD~\cite{pat}
describes the $pp$ data used as a reference in the experimental studies.
This is important since, for the LHC measurement, the $pp$ data are
estimated by using a variety of scaling properties of the particle production
in hadronic collisions.

In perturbative QCD, the inclusive cross-section for the production
of particle $h$ in
a nucleon-nucleon  collision is given by:
\begin{eqnarray}
E_h\frac{d^3\sigma}{d^3p_h}(\sqrt{s},p_h)
 =\int \, dx_a \int \, dx_b \int \, dz \sum_{i,j} & & F_i(x_a,Q^2)
F_j(x_b,Q^2)\nonumber\\
& & D_{c/h}(z,Q^2_f)E_c\frac{d^3\sigma_{ij \rightarrow cX}}{d^3p_c}~,
\end{eqnarray}
where $F_i(x,Q^2)$ is the distribution function for the $i$-th parton in a
nucleon, $D_{c/h}$ is hadron fragmentation function at $z=p_h/p_c$, and
 $\sigma_{ij \rightarrow cX}$ is parton-parton cross-section. We include leading
order, ${\cal{O}}(\alpha_s^2)$ processes, like:
\begin{eqnarray}
q + q &\rightarrow &q +q , \nonumber\\
q+\bar{q} &\rightarrow& q+\bar{q} ,\nonumber\\
q+g &\rightarrow& q+g,\nonumber\\
g+g &\rightarrow & g+g,\nonumber\\
....
\end{eqnarray}
and the next-to-leading order, ${\cal{O}}(\alpha_s^3)$, sub-processes
such as:
\begin{eqnarray}
q+q \rightarrow q+q+g~,\nonumber\\
q+\bar{q} \rightarrow q+\bar{q}+g,\nonumber\\
q+q^{\prime} \rightarrow q+q^{\prime}+g, \nonumber\\
q+\bar{q} \rightarrow q^{\prime}+\bar{q}^{\prime}+g, \nonumber\\
g+g \rightarrow g + g+ g\nonumber\\
....
\end{eqnarray}
The running coupling constant $\alpha_s(\mu^2)$, calculated at next-to-leading
order, is given by
\begin{equation}
\alpha_s(\mu^2)=\frac{12\pi}{(33-2N_f)\ln(\mu^2/\Lambda^2)}
\left[1-\frac{6(153-19N_f)\ln \ln (\mu^2/\Lambda^2)}
{{(33-2N_f)}^2\ln(\mu^2/\Lambda^2)}\right]
\end{equation}
where $\mu$ is the renormalization scale, $N_f$ is number of flavours, and
$\Lambda=\Lambda_{\textrm{QCD}}$. We use CTEQ4M structure functions~\cite{cteq}
and Binnewies, Kniehl, and Kramer~\cite{bkk} fragmentation functions. We use 
the factorization, renormalization, and fragmentation scales as 
$Q=p_T$, though we have checked the results for the particle production in $pp$
collisions using scales $Q=p_T/2$ and $Q=2p_T$ as well. 

Our results along with the $\pi^0$ data measured by the PHENIX
experiment~\cite{phenix_pp}
are shown in Fig.~\ref{fig1}. 
We see a very good  description of the experimental results with-out any
free parameters. Similar results have been repeatedly reported earlier
(see e.g. Ref.~\cite{phenix_pp}).  

For the $pp$ centre of mass energy of 2.76 TeV, no experimental data is 
available and the ALICE collaboration has utilized a multi-pronged approach to
estimate~\cite{jet_lhc}
 this using several well established scaling relations. This estimate
and our calculations are shown in Fig.~\ref{fig2}. The good agreement
of the estimated cross-sections with the NLO pQCD results, holds out the
hope of getting an accurate measure of the medium modification of
$R_\textrm{AA}$ in such studies.
\begin{figure}
\begin{center}
\includegraphics[width=23pc,angle=-0]{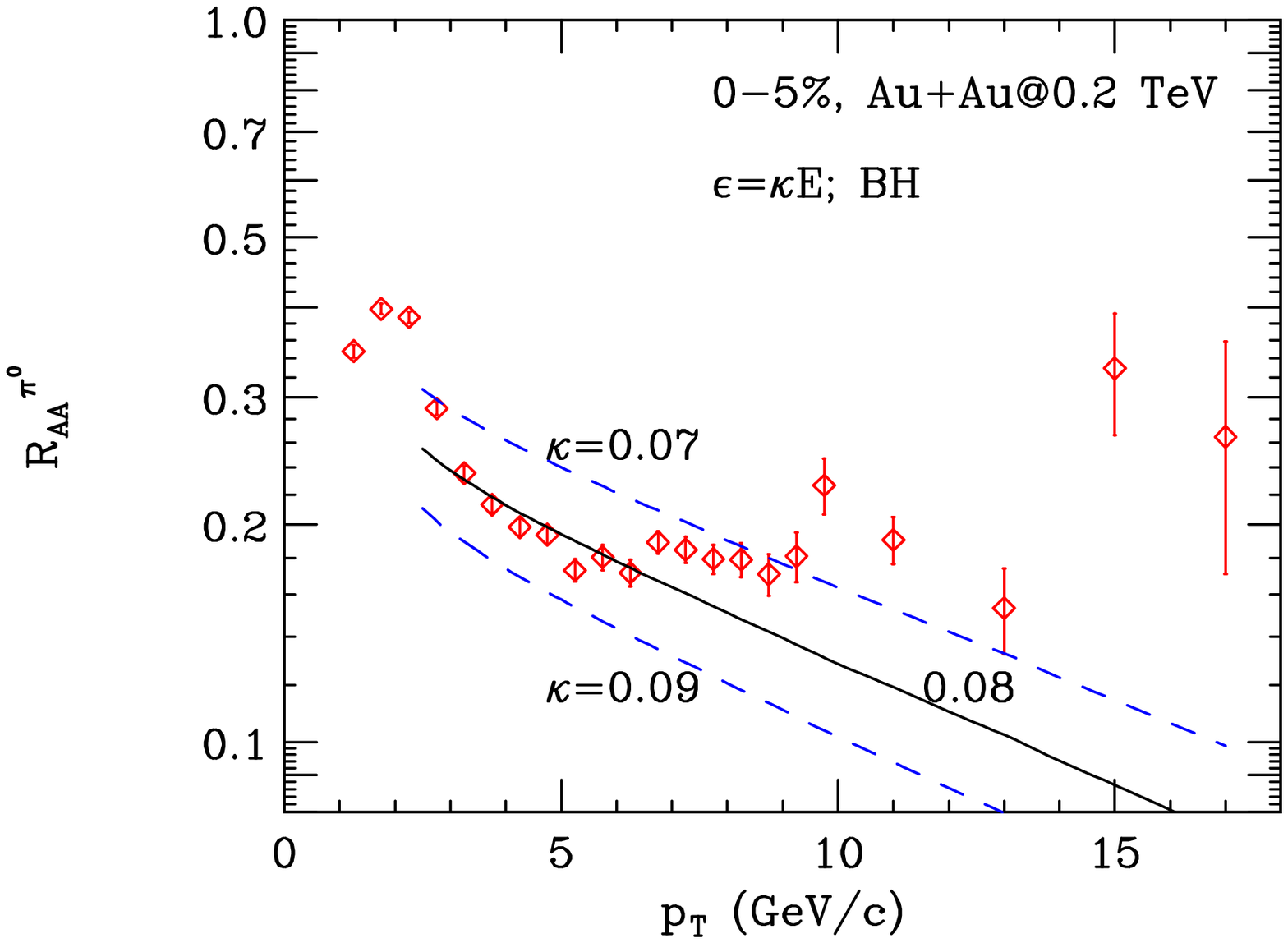}
\includegraphics[width=23pc,angle=-0]{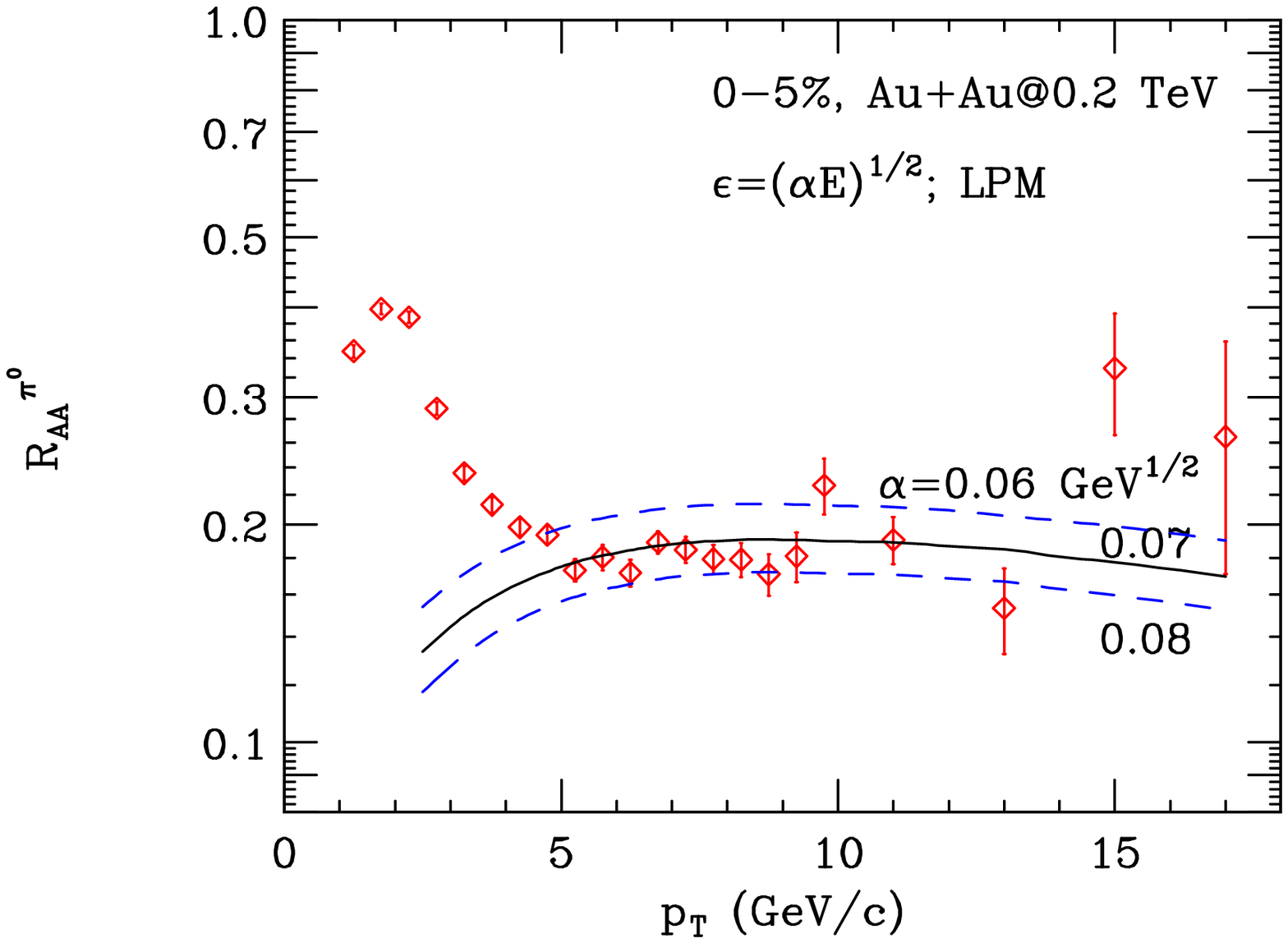}
\includegraphics[width=23pc,angle=-0]{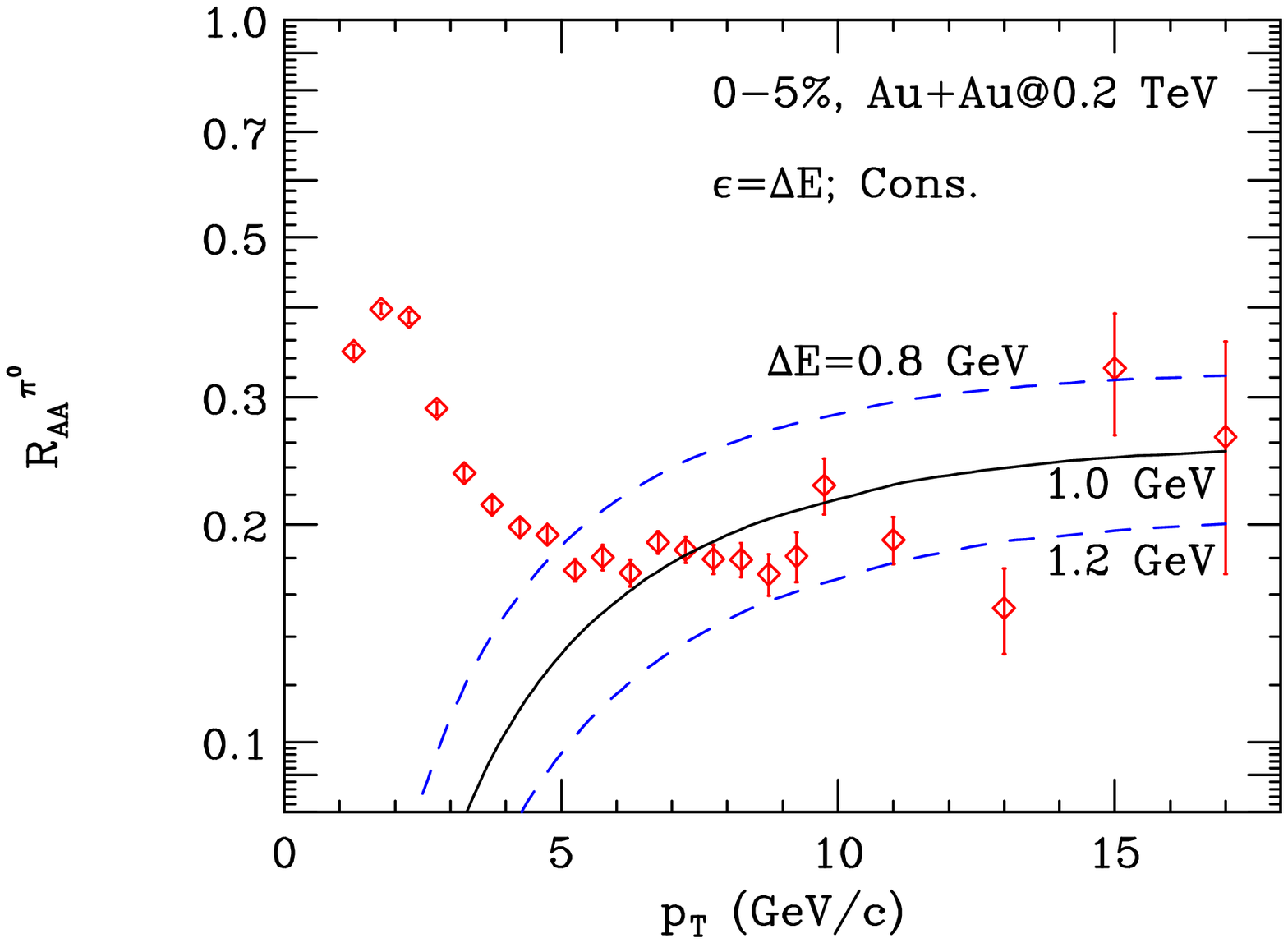}
\caption{The nuclear modification function for the production of
neutral pions for the 0--5\% most central
collision of gold nuclei at the top RHIC energy. $\epsilon$ is
energy loss per collision and the
mean free path is taken as 1 fm.  The 
experimental results are from the PHENIX collaboration~\cite{phenix_pi0}.
}
\label{fig3}
\end{center}
\end{figure}

\subsection{Medium modification of $R_\textrm{AA}$}

By now it is very well established that the medium modification of
 $R_\textrm{AA}$ is  
mainly caused by the energy-loss suffered by partons while they traverse the 
plasma due to collisions and radiation of gluons, before they fragment and
the nuclear shadowing.  The nuclear shadowing accounts for the
modification of distribution of quarks and gluons inside nuclei
compared to that in free nucleons. We use the
nuclear shadowing function obtained by Eskola, Kolhinen, 
and Salgado~\cite{eks98} which is known to describe the NMC data accurately.

A substantial body of literature exists on the dynamics of 
propagation of fast partons through the hot and dense medium produced in
relativistic heavy ion collisions, where they lose part of their
energy through collisions and radiation of gluons before 
fragmenting into hadrons. 

These developments can be broadly categorized as follows:

\begin{itemize}

\item The Higher Twist Approximation developed by Guo and Wang~\cite{HT}, which
is based upon the huger twist correction for final state partons in 
$e+A$ collisions. 

\item The so-called AMY formalism developed by Arnold, Moore, 
and Yaffe~\cite{AMY} based on hard thermal loop resummation in a pertubative
plasma.

\item The so-called ASW formalism developed by Armesto, Salgado, 
and Wiedemann~\cite{ASW} which resums multiple soft gluon emission in the 
BDMPS approach~\cite{baier} (see later) in a finite length medium using 
Poisson statistics, and

\item The GLV formalism developed by Gyulassy, Levai, and Vitev~\cite{GLV}
      which considers scattering centres in the opacity expansion.

\end{itemize}

Several developments tracing the evolution of the entire jet in the
medium using Monte Carlo jet-quenching modules, e.g. PYQUEN~\cite{pyquen},
Q-PYTHIA~\cite{q-pythia}, JEWEL~\cite{jewel}, YaJEM~\cite{yajem}, 
and MARTINI~\cite{martini} have also been reported. Many of these developments
have been discussed in detailed reviews~\cite{rainer,abhijit}.

We use a simple phenomenological model~\cite{wang} to study the 
evolution of the mechanism of energy loss with increasing $p_T$
and centre of mass energy. This has been shown to be quite successful
in explaining the results for $R_\textrm{AA}$~\cite{jeon} at RHIC energies.
It has also been used extensively to illustrate various consequences of
jet-quenching, e.g. photon-tagged jets and di-hadron correlations etc. 

The advantage of this treatment, as we shall see, lies in giving reasonable
description of the medium modification in terms of energy loss per collision
and the mean-free of the parton, which covers an average path length 
in the medium. We shall see that this works quite well and also helps us
to distinguish the energy dependence of the energy loss mechanism.

We focus our attention on the central rapidity and central collisions,
so that the parton propagates in the transverse direction and the azimuthal
dependence of the path-length etc. can be safely ignored. Taking the path
length as $L$ and the mean free-path of the parton $a$ as $\lambda_a$, we 
estimate the probability for a parton to scatter $n$ times before escaping
as
\begin{equation}
P_a(n,L)=\frac{(L/\lambda_a)^n}{n!}\,e^{-L/\lambda_a}~.
\end{equation}
 Following Ref.~\cite{wang}, we modify the hadronic fragmentation function
$D_{c/h}(z,Q^2)$ to include multiple scattering and the energy loss
of the parton in the nuclear medium. Assuming that the average energy loss
per collision suffered by the parton $a$ is $\epsilon_a$, the 
nuclear fragmentation function can be written as:
\begin{eqnarray}
zD_{c/h}(z,L,Q^2)=\frac{1}{C_N^a}
\sum_{n=0}^N \, 
 P_a(n,L)& & \left[\, z_n^a \, D_{c/h}^0(z_n^a,Q^2)\right.\nonumber\\
& &\left. +\sum_{m=1}^n \,z_m^a\,D_{g/h}^0(z_m^a,Q^2)\right]~,
\label{mod_frag}
\end{eqnarray}
where $z_n^a=zE_T^a/E_n^a$, $E_{n+1}^a=E_n^a-\epsilon_n^a$,
$z_m^a=p_T/\epsilon_m^a$, $n$ 
is the maximum number of collisions for which $z_n^a \leq 1$,  $D_{c/h}^0$
is the hadronic fragmentation function which gives the probability for
the fragmentation of a quark or a gluon to fragment into a $\pi^0$ or
a charged particle (or any particular hadron in general), and
\begin{equation}
C_N^a= \sum_{n=0}^N \, P_a(n,L)~.
\end{equation}
Of-course $E_0^a=E_T^a$ in an obvious notation.
 In the Eq.~\ref{mod_frag}, the 
first term corresponds to the fragmentation of the leading parton and the
second term comes from the emitted gluons having energy $\epsilon_a^m$. 
The average number of scatterings over the distance $L$ is $L/\lambda_a$,
where we take $L=1.2(0.5\times N_\textrm{part})^{1/3}$ and $\lambda_a$ =1 fm.
(This value of $L$  should suffice for the most central collisions considered
here. For large impact-parameters $L$ will vary considerably with 
the azimuthal angle and its average value will be smaller.
Consequences of this variation along with those of
 changing $\epsilon_a$ and $\lambda_a$ with 
quarks/gluons will be reported shortly.) As mentioned earlier,
we use BKK fragmentation functions from ref.~\cite{bkk} and take the
factorization, renormalization, and fragmentation scales to be equal to $p_T$.
\begin{figure}
\begin{center}
\includegraphics[width=23pc,angle=-0]{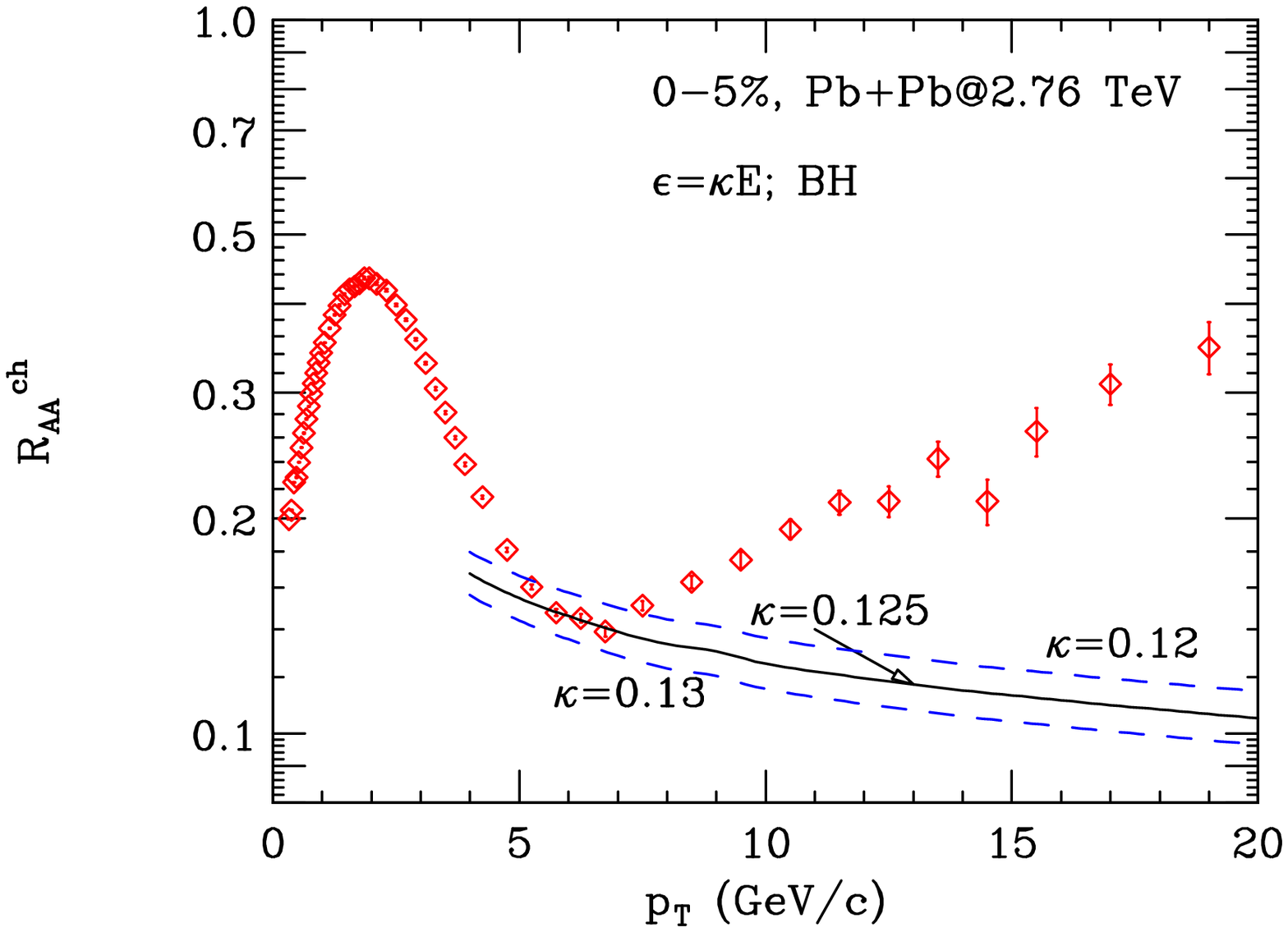}
\includegraphics[width=23pc,angle=-0]{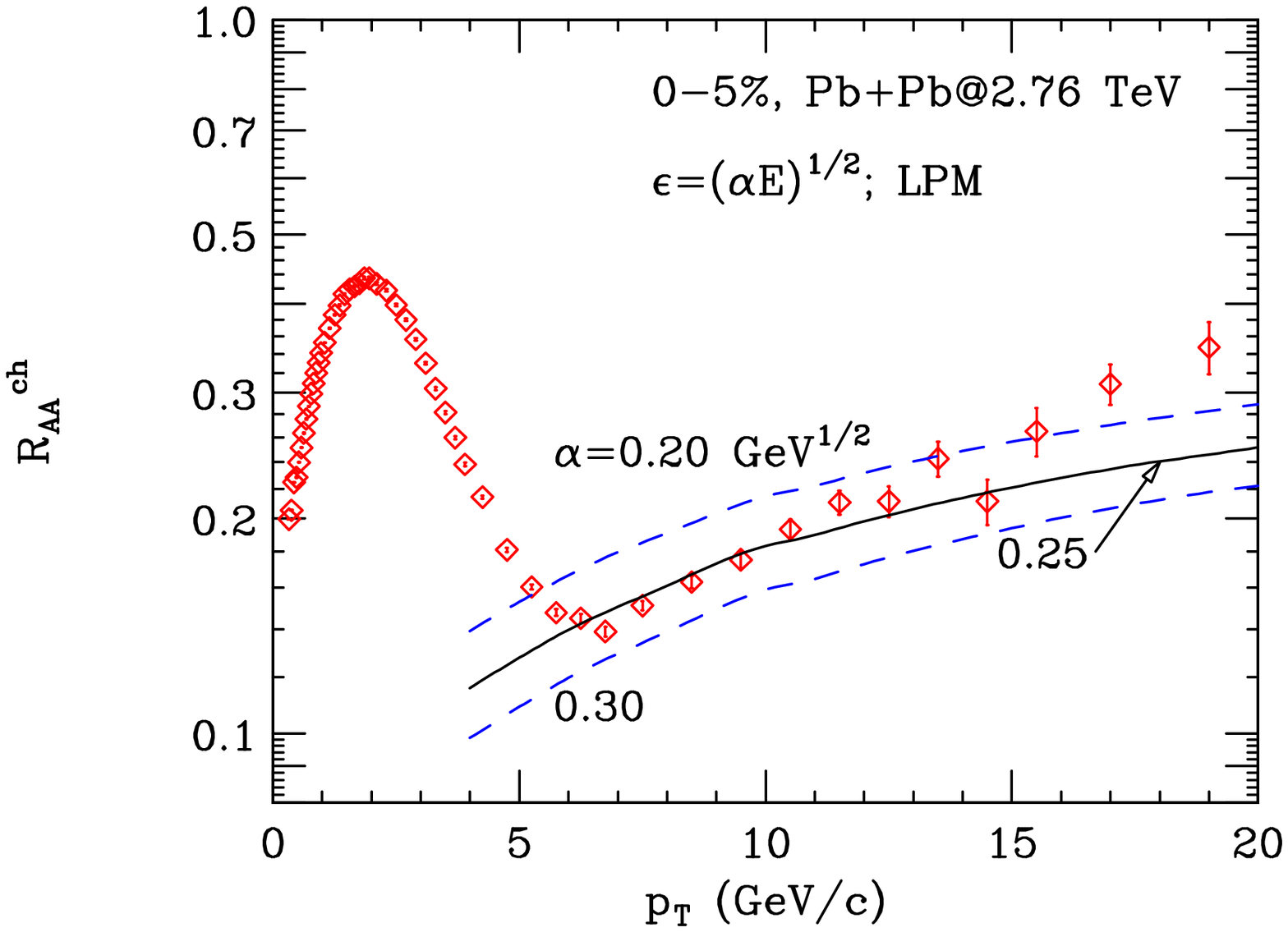}
\includegraphics[width=23pc,angle=-0]{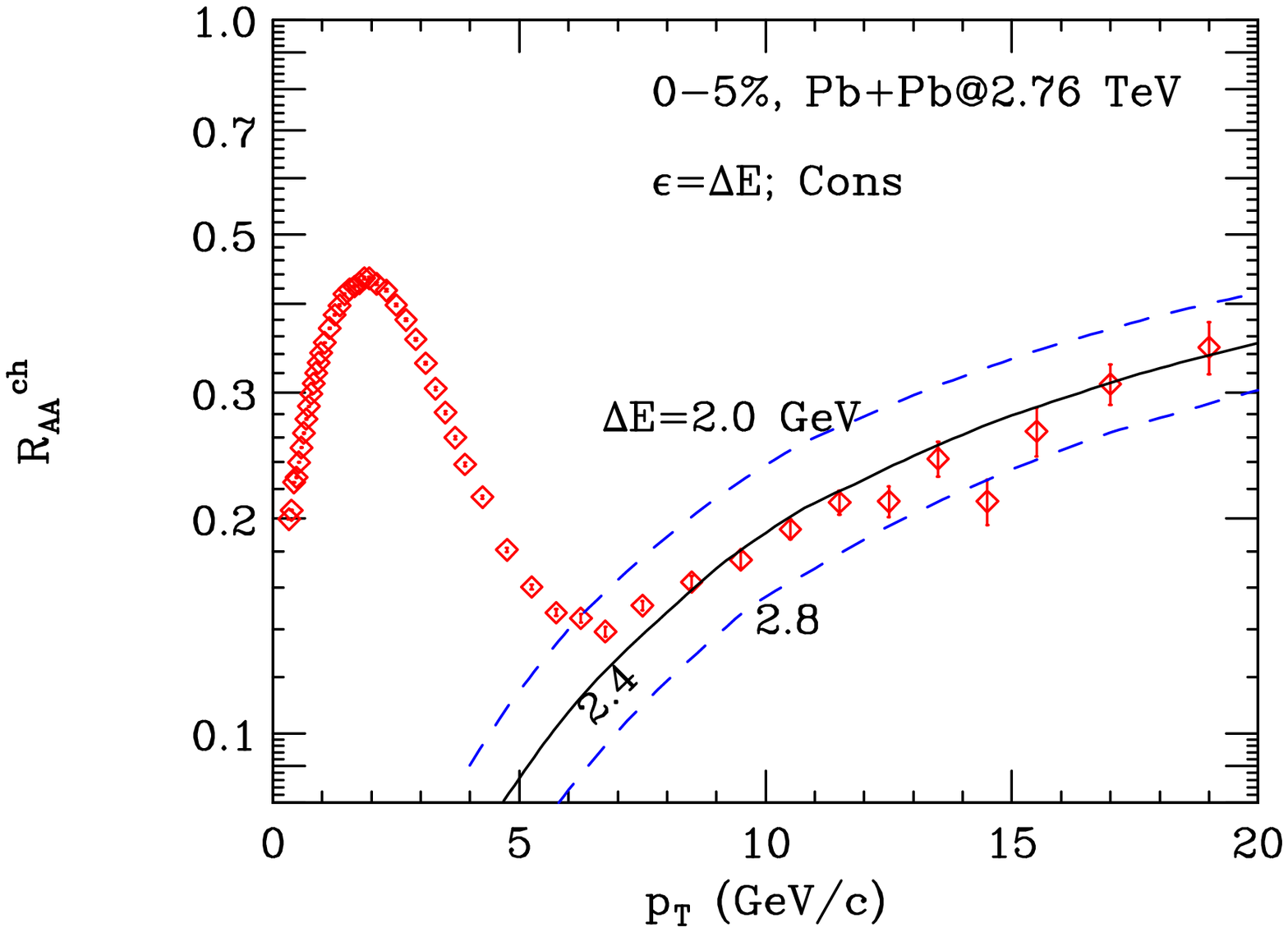}
\end{center}
\caption{The nuclear modification function for the production of
charged particles for the 0--5\% most central
collision of lead nuclei at $\sqrt{s_\textrm{NN}}$= 2.76 TeV
measured at the Large Hadron Collider. $\epsilon$ is energy loss
per collision and the mean free path
is taken as 1 fm.  The experimental data 
are from the ALICE experiment~\cite{jet_lhc}.}
\label{fig4}
\end{figure}
\subsection{Parton energy loss}

We employ three different prescriptions for the the energy loss 
per collision ($\epsilon_a=\lambda_a\times dE_a/dx$). 
The choices are inspired by the excellent discussion in
Baier et. al~\cite{baier}. These authors argue that for light quarks and gluons
the collisional energy loss ($dE/dx$) is quite small ( 0.2 -- 0.3 GeV/fm for 
quarks and $\frac{9}{4}$ times this value for gluons at $T\approx$~ 250 MeV).
 Thus the
energy loss for them is dominated by the mechanism of radiation of gluons. 

The radiation of gluons is then best discussed in terms of the formation
time of the radiated gluon:
\begin{equation}
t_\textrm{form}\approx \frac{\omega}{k_T^2}~,
\end{equation}
where $\omega \gg k_T$ is the energy of the gluon and $k_T$ is the
transverse momentum of the gluon.
 For small $\omega$ ($t_\textrm{form} \leq \lambda$) incoherent 
radiation takes place over $L/\lambda$ scattering centres. This is the 
so-called Bethe-Heitler regime and one can derive:
\begin{equation}
-\frac{dE}{dx}\approx\frac{\alpha_s}{\pi}N_c\frac{1}{\lambda}E~
\label{BH}
\end{equation}
where $N_c=3$. We shall write $\epsilon \approx \kappa E$ for this case 
and determine $\kappa$ using the $R_\textrm{AA}$ measured in the
experiment.
\begin{figure}
\begin{center}
\includegraphics[width=25pc,angle=0]{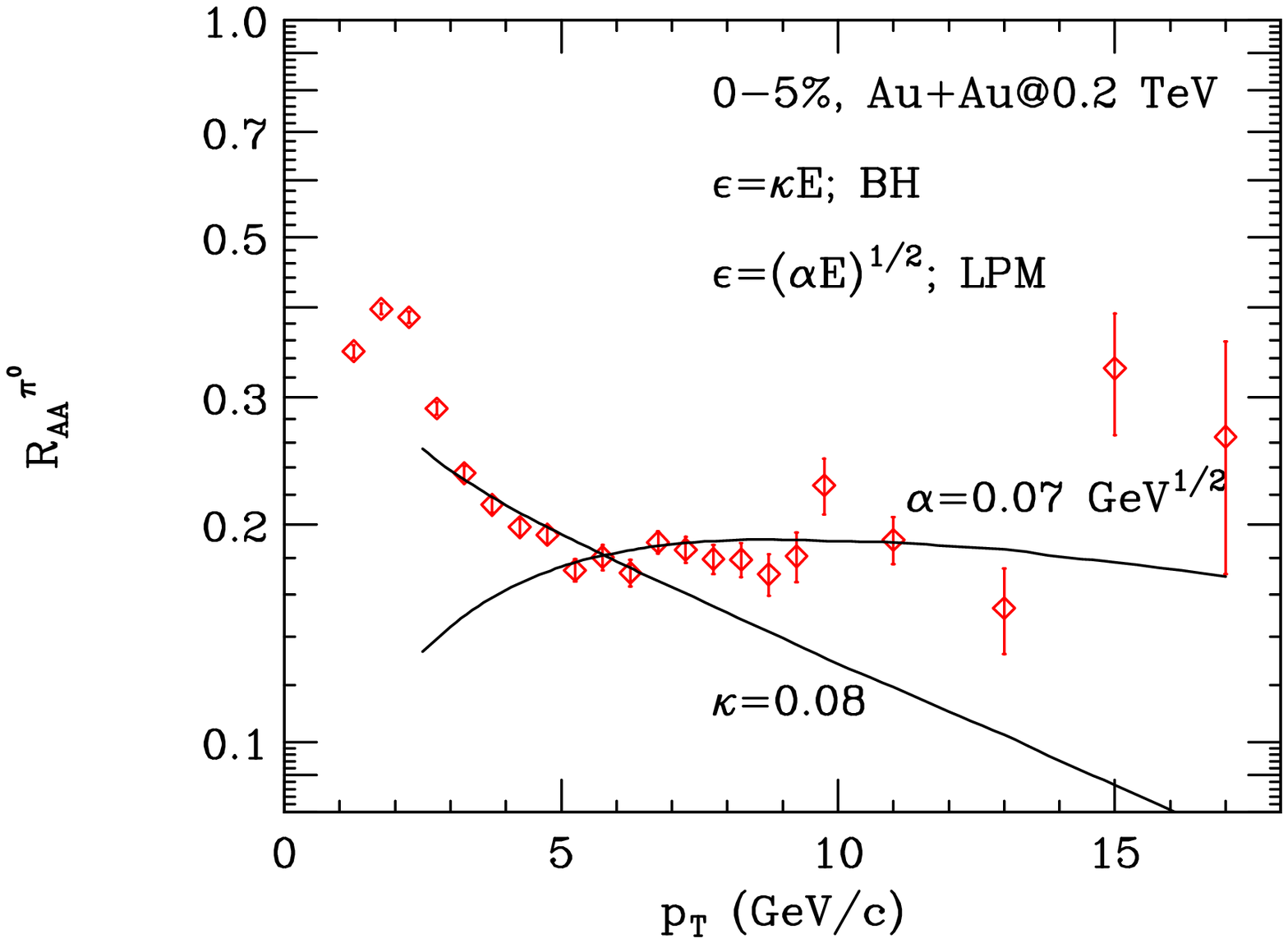}
\includegraphics[width=25pc,angle=0]{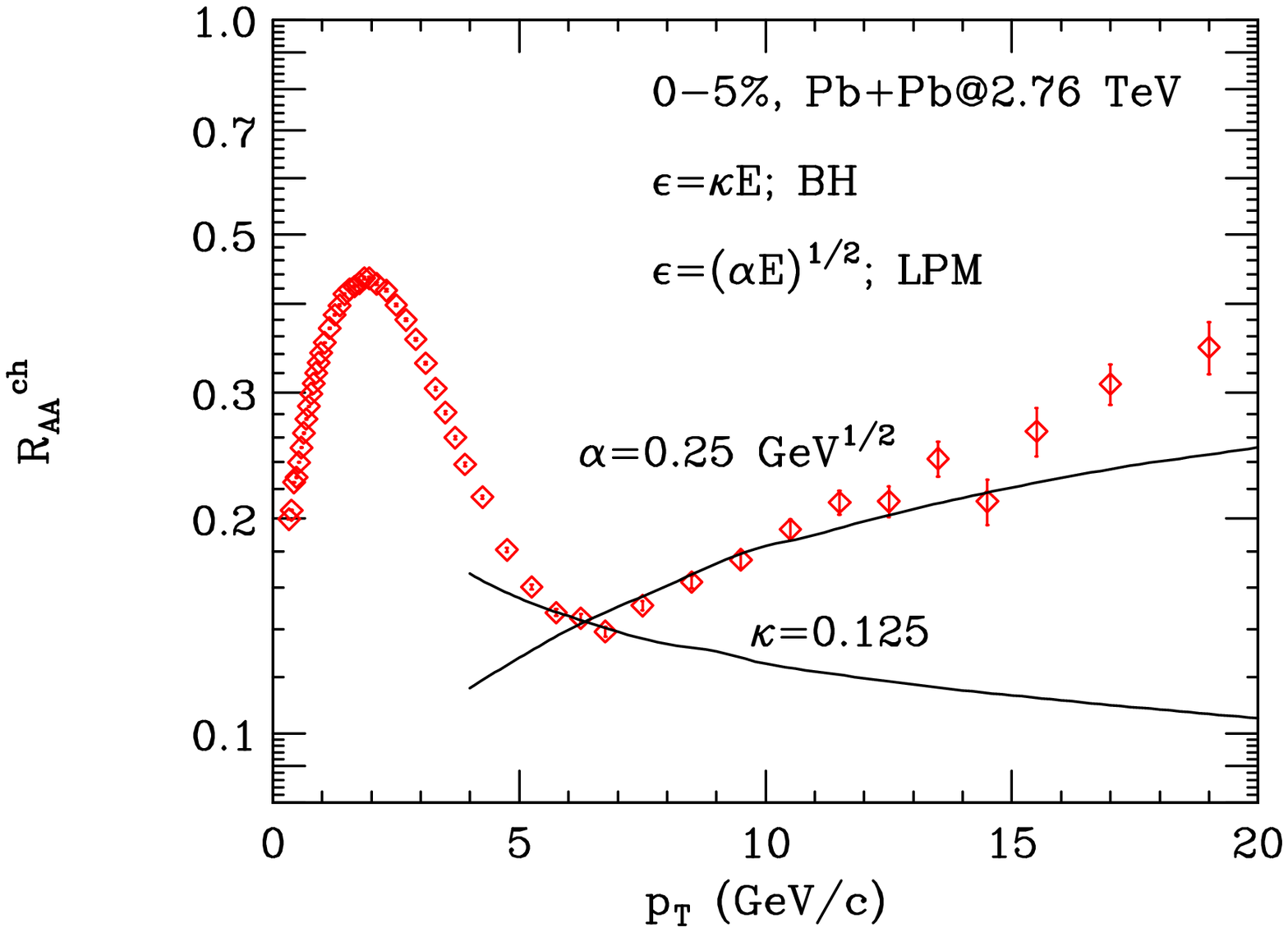}
\end{center}
\caption{Evolution of the mechanism of energy loss from Bethe Heitler
to Landau Pomeranchuk Midgal as a function of $p_T$ at RHIC and LHC.}
\label{fig5}
\end{figure}

When the formation time $t_\textrm{form}$ is greater than $\lambda$ but
less than $L$, $N_\textrm{coh}$ scattering centres act as a 
single source of radiation, where
\begin{equation}
N_\textrm{coh}=\ell_\textrm{coh}/\lambda
\end{equation}
and
\begin{equation}
\ell_\textrm{coh}\approx \frac{\omega}{<k_T^2>_\textrm{coh}},
\end{equation}
where $<k_T^2>_\textrm{coh}$ is the accumulated transverse momentum
and $\approx N_\textrm{coh}\times <k_T^2>$. Thus
one can write:
\begin{equation}
N_\textrm{coh}\approx\sqrt{\frac{\omega} {\lambda \,<k_T^2>}}=
\sqrt{\frac{\omega}{E_\textrm{LPM}}}
\end{equation}
where $E_\textrm{LPM}=\lambda <k_T^2>$ is the energy which separates
the BH regime from the LPM regime. Now, one can write:
\begin{equation}
-\frac{dE}{dx}\approx \frac{\alpha_s}{\pi}
\frac{N_c}{\lambda} \sqrt{E_\textrm{LPM} E}~.
\label{LPM}
\end{equation}

In the present work, we denote this scenario with 
$\epsilon=\sqrt{\alpha E}$ and
determine $\alpha$ from a description of $R_\textrm{AA}$.

Finally, when the formation time is much larger than the path-length $L$
the coherence is complete and
the medium induced total energy loss becomes proportional to the square 
of $L$, and one gets:
\begin{equation}
-\frac{dE}{dx}=\frac{\alpha_s}{\pi}N_c \frac{<k_T^2>}{\lambda} L~.
\label{L}
\end{equation}

We shall denote this case by having $\epsilon$ = constant, for a
given $L$. One may need to check it for cases
having varying $L$ (e.g., centrality dependence of $R_\textrm{AA}$). Results
for such a study would be published shortly.
\begin{figure}
\begin{center}
\includegraphics[width=25pc,angle=0]{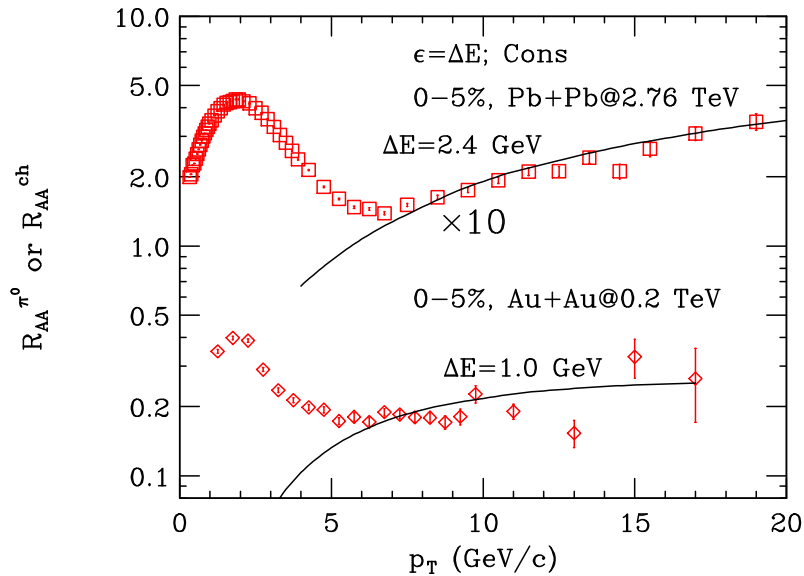}
\end{center}
\caption{Doubling of energy loss per collision as  
$\sqrt{s_\textrm{NN}}$ is increased from 200 GeV to 2.76 TeV.}
\label{fig6}
\end{figure}
The NLO code of Aurenche et al had earlier been modified by
Jeon et al~\cite{jeon} to incorporate the medium modified
 fragmentation functions.
We use it for the calculations reported here.  

\section{Results}

We have already seen that the NLO pQCD calculations accurately describe the
results for particle production in $pp$ collisions at both the energies
under consideration (See Figs.~\ref{fig1},\ref{fig2}).

 The nuclear modification factor $R_\textrm{AA}$ obtained in the 
present work for central collision of gold nuclei at $\sqrt{s_\textrm{NN}}$
equal to 200 GeV, using the three energy loss mechanisms discussed
above are shown in Fig.~\ref{fig3}. The so-called Bethe Heitler mechanism
is seen to accurately describe the nuclear modification for 
2.5~$<~p_T~<$~6 GeV/$c$ (top panel).

 It is interesting to note that the 
condition necessary for the multiple scatterings to be incoherent,
when the mean free path is about 1 fm requires that the formation
time of the gluon is less than 1 fm/$c$. This will be satisfied if
the energy of the radiated gluon is less than 5 GeV and 
$<k_T^2> \approx$ 1 $(\textrm{GeV}/$c$)^2$.  Thus 
indeed the partons having energy up to 5--7 GeV would populate this
 $p_T$ window. The value of $\kappa \approx$ 8\%, obtained here
is close to (but higher than) that obtained by
Jeon et al.~\cite{jeon} for 0--10\% centrality. However it is smaller
than the value suggested by the Eq.~\ref{BH}. We shall discuss this later.

The nuclear modification shows a clear change in slope around $p_T$ equal to
5 GeV/$c$. We also note that $E_\textrm{LPM}\approx$ 5 GeV as discussed
above.  Thus it is interesting 
that the energy loss implied by Eq.~\ref{LPM} provides an accurate
description of $R_\textrm{AA}$ for 5 GeV/$c$ $< p_T < $ 12 GeV/$c$,
 when we take $\alpha$ as 0.07 GeV$^{1/2}$ (Fig.~\ref{fig3}, middle
panel). We again see that while the form
of the energy loss is given by Eq.~\ref{LPM}, the coefficient for the
energy loss we get is about half as small.

The statistics for the modification factor is quite poor at larger $p_T$,
however the average behaviour seen (see Fig.~3, lower panel) is in fair
agreement with energy loss mechanism given by Eq.~\ref{L}, 
though once again the energy loss per collision is smaller than that
implied by this equation.

The results for $\sqrt{s_\textrm{NN}}$ = 2.76 TeV, (see Fig.~\ref{fig4})
are also quite interesting. We now see (top panel) that the BH mechanism
provides a description of the data over a limited $p_T$ 
window of 5 -- 7 GeV/$c$.
It is quite likely that at these higher energies the validity of hydrodynamics
may extend to higher $p_T$ than that witnessed at RHIC energies and thus
the radial flow may affect the $p_T$ window up to 5 GeV/$c$ 
as well as our results.

The so-named LPM regime is seen to hold over the $p_T$ window of 7--14 GeV/$c$
(see Fig.\ref{fig4}, middle panel) 
with $\alpha \approx$ 0.25 GeV$^{1/2}$, which is still smaller (by about 30\%)
 than that implied by Eq.~\ref{LPM}. 

The mechanism which gives a constant energy loss per collision is seen
to accurately describe the data over the largest $p_T$ window of 10 -- 20
GeV/$c$ (see Fig.\ref{fig4}, lower panel). This is interesting as 
this mechanism is supposed to work for $L < L_\textrm{cr}$ 
where $L_\textrm{cr}=\lambda \sqrt{E/E_\textrm{LPM}}$ which should be true
for $ E \geq$ about 125 GeV for $L=$ 5 fm, while it 
seems to be operating even at lower energies. 
The energy loss per collision is still
smaller than that implied by Eq.~\ref{L}, though it is about twice as large
as at the top RHIC energy. 

In order to bring out the changing mechanism for energy loss
with increasing $p_T$ we now  plot the final results for RHIC and
LHC energies using the BH and LPM mechanisms (see Fig.~\ref{fig5}, upper and
lower panels). The results at RHIC energies show a dramatic change in
the mechanism around  $p_T$ of 5 GeV/$c$, as indeed one expects from the
consideration of formation time of the radiated
gluons and the mean free path of the partons.

The comparison for the constant energy loss per collision for the RHIC
and the LHC energies is shown in Fig.~\ref{fig6}. We see that while the
results at RHIC are only indicative of possible emergence of this 
mechanism at larger $p_T$, the results at LHC clearly favour this 
mechanism for $p_T >$ 10 GeV/$c$.

\section{Summary and Discussions}

We have adopted a simple and transparent model of energy loss of partons
traversing the medium created in relativistic collision of heavy nuclei.
The fragmentation functions of the partons are then modified to account for the
energy loss (and also fragmentation of radiated gluons) and nuclear
modification function $R_\textrm{AA}$ calculated. The energy loss per
collision is taken to be either proportional to the energy of the
parton, square-root of the energy of the parton, or a constant, inspired by
the treatment of energy loss depending on the formation time of the radiated
gluon and the mean free path of the parton. The reference particle spectra 
are calculated using NLO pQCD.

A dramatic change in the energy loss mechanism is seen at $p_T \approx$ 
5 GeV/$c$.  We find that the Bethe- Heitler mechanism of incoherent
 scattering prevails
for $p_T \leq$ 5 GeV/$c$ at both RHIC and LHC energies, while the
so-called LPM mechanism, which gives energy loss per collision as proportional
to the square-root of the energy prevails over the $p_T$ window of
 5--15 GeV/$c$. In an interesting observation we find that the constant 
energy loss per collision suggestive of a total energy loss increasing with
the square of the path length may already be becoming relevant for $p_T >$
10 GeV/$c$. This mechanism is seen to prevail over the $p_T$ window of
10--20 GeV/$c$ at $\sqrt{s_\textrm{NN}}$ = 2.76 TeV. 

While the results generally follow the qualitative arguments given in
Ref.~\cite{baier} for the form of the energy-loss per collision as the
energy of the parton changes, the energy-loss coefficients are 
smaller than those implied in the above work. One obvious difference is 
while the medium produced in these collisions is expanding and cooling
the treatment of Ref.~\cite{baier} is basically for a static medium 
(with static centres of scattering, a la Glauber model), though modifications
 for expansion are available. Obviously, in an expanding medium
the parton will initially move in a hotter layer and later in cooler
layers. The energy loss per collision emerging in the present work 
will then give an average over the history of evolution. The expansion
of the plasma will also lead to a $L$ larger than the radius of the
system, at least at LHC energies. However the expanding volume beyond the
volume of the plasma produced originally will also be cooler 
and thus the energy loss there may be small. It is also likely that the
partons produced near the centre would have already lost most of their
energy by the time they enter this volume, and thus they would
fragment to low energy hadrons. 
 The boundaries of the different mechanisms may
also not be quite sharp.

It will be interesting to see how these mechanisms evolve with the
centrality of the collision~\cite{ldep} and how do they fare
when the dynamics of evolution is accounted for~\cite{steffen}. 
Is this simple but transparent treatment able to account for the
azimuthal anisotropy of the momentum distribution for non-central
collisions at large $P_T$? These will be addressed in a forthcoming 
publication ~\cite{somnath}. We add that, even though the discussion
of Ref.~\cite{baier} focuses on radiative energy loss of the partons 
the treatment in terms of energy loss per collision could easily 
accommodate the collisional energy loss, as long as it is small.

We conclude that the mechanism of energy loss of the partons weaves
a rich tapestry depending on the energy of the parton and 
the properties of the medium. The simple model of multiple scattering
used in the work is able to reveal the evolution of the 
mechanism of energy loss with $p_T$.

\section*{Acknowledgments} 
Valuable discussions with Somnath De, Umme Jamil, Sangyong Jeon, and
Mohammed Younus are gratefully acknowledged.

\end{document}